\input{epsf}

\documentclass[conference]{IEEEtran}

\makeatletter
\def\ps@headings{%
\def\@oddhead{\mbox{}\scriptsize\rightmark \hfil \thepage}%
\def\@evenhead{\scriptsize\thepage \hfil \leftmark\mbox{}}%
\def\@oddfoot{}%
\def\@evenfoot{}}
\makeatother
\usepackage{epsf}
\usepackage{graphicx}
\usepackage[cmex10]{amsmath}
\usepackage{amsthm}
\usepackage{amssymb}
\usepackage{epsfig,latexsym,amsmath,epsf,amssymb,amsfonts}
\usepackage{algorithm, algorithmic}
\usepackage{placeins}
\usepackage{url}
\usepackage{cite}
\usepackage{comment}
\usepackage{subfigure}
\usepackage{lipsum}
\usepackage{physics}
\usepackage{color}
\usepackage{tikz}
\usetikzlibrary{trees, shapes, positioning, arrows, arrows.meta}

\usepackage[top=0.75in, bottom=1in, left=0.625in, right=0.625in]{geometry}
\usepackage{bbm}
\usepackage{url}

\newtheorem{Definition}{\hskip 0pt Definition}
\newtheorem{Theorem}{\hskip 0pt Theorem}
\newtheorem{Corollary}{\hskip 0pt Corollary}

\begin{document}
\title{Decentralized Caching for Content Delivery Based on Blockchain: A Game Theoretic Perspective}
\author{
\IEEEauthorblockN{Wenbo Wang\IEEEauthorrefmark{1},
Dusit Niyato\IEEEauthorrefmark{1},
Ping Wang\IEEEauthorrefmark{1}
and
Amir Leshem\IEEEauthorrefmark{2}}
\IEEEauthorblockA{\IEEEauthorrefmark{1}School of Computer Engineering, Nanyang Technological University, Singapore 639798}
\IEEEauthorblockA{\IEEEauthorrefmark{2}Faculty of Engineering, Bar-Ilan University, Ramat-Gan, Israel, 52900}\vspace*{-8mm}}
\maketitle
\begin{abstract}
  Blockchains enables tamper-proof, ordered logging for transactional data in a decentralized manner over open-access, overlay peer-to-peer networks. In this paper, we propose a decentralized framework of proactive caching in a hierarchical wireless network based on blockchains. We employ the blockchain-based smart contracts to construct an autonomous content caching market. In the market, the cache helpers are able to autonomously adapt their caching strategies according to the market statistics obtained from the blockchain, and the truthfulness of trustless nodes are financially enforced by smart contract terms. Further, we propose an incentive-compatible consensus mechanism based on proof-of-stake to financially encourage the cache helpers to stay active in service. We model the interaction between the cache helpers and the content providers as a Chinese restaurant game. Based on the theoretical analysis regarding the Nash equilibrium of the game, we propose a decentralized strategy-searching algorithm using sequential best response. The simulation results demonstrate both the efficiency and reliability of the proposed equilibrium searching algorithm.
\end{abstract}

\section{Introduction}\label{Sec_introduction}
According to the latest Cisco VNI report~\cite{index2016global}, over the past 5 years the global mobile data traffic has grown 18-fold, which is largely driven by the explosive growth of the on-demand mobile video traffic. In contrast, it is predicted that the mobile network connection speed will only increase 3-fold in the next 5 years. Motivated by the fact that most mobile traffics are asynchronous but repetitive content delivery requests, the research communities have vastly investigated the technique of proactive caching in recent years~\cite{7931641,7283673,7797148}. In general, proactive caching consists of two stages. The first stage is the prefetching stage, where the edge servers such as Wi-Fi Access Points (APs) and Small-Cell Base Stations (SCBSs), or even Device-to-Device (D2D)-enabled nodes pre-download/replicate the contents from the Content Servers (CSs) before the content requests are posted by any mobile users. The second stage is the delivery stage, where the contents are delivered to the mobile users via the fronthaul or D2D links from the caching nodes. A recent line of works has shown that proactive caching is able to relieve the backhaul congestion~\cite{7931641}, reduce the cost of the content providers~\cite{7283673} and/or improve the Quality of Experience (QoE) of the mobile users~\cite{7797148}.

With the goals of offloading the heavy traffic and ensuring delivery in case of high volatility, proactive caching at the APs/SCBSs in wireless networks shares a lot of similarities with the commercial Content Delivery Networks (CDNs)~\cite{Pathan06ataxonomy}. However, due to the limited resource in storage, computation power and edge-to-end capacity, an AP/SCBS can only afford caching a subset of the contents from the Content Providers (CPs). Then, how to properly select the contents for proactive caching becomes a vital challenge for the edge nodes, namely, Cache Helpers (CHs). Moreover, due to the self-deployment nature of the CHs and the coexistence of multiple CPs, it is impractical to deploy or coordinate the accounting/broking servers which are required by the logical system of a traditional CDN~\cite{Pathan06ataxonomy} for transaction auditing and traffic estimation. As a result, no trusted entity in the network audits the content access/delivery or enforces the proper payments to the right parties (e.g., CPs and CHs). Therefore, new expectations are set for a decentralized, self-organized proactive caching system:
\begin{itemize}
  \item [(1)] The brokering and accounting processes between the CPs and CHs should be achieved in a decentralized market, with truthfulness being ensured for both parties.
  \item [(2)] The CHs should be able to adapt to the changes of content demand/supply in the network without access to a centralized content delivery accounting server.
  \item [(3)] The processes of cache selection and delivery allocation should remain transparent to the mobile users.
  \item [(4)] Proper incentive mechanisms should be provided to sustain self-motivated content caching among the APs/SCBSs.
\end{itemize}

In this paper, we jointly address the challenges above by constructing an autonomous content-caching market with a smart contract-enabled blockchain. The blockchain serves as a public immutable ledger without any centralized mediator in a logical Peer-to-Peer (P2P) network~\cite{7423672}. The transaction records between blockchain users are jointly approved by the consensus nodes and digitally stored in their local blockchain replicas. With the blockchain, the content prefetching and delivering processes are self-organized in the form of smart contracts~\cite{wood2014ethereum}. The blockchain also provides publicly accessible records about the demand and supply of contents in the network. By jointly considering the CHs' activities in smart contract execution and block consensus maintenance, we adopt a cross-layer design for the blockchain's consensus mechanism based on Proof-of-Stake (PoS)~\cite{Kiayias2017}. Compared with the existing centralized caching systems, the key advantages of the proposed mechanism are:
\begin{itemize}
 \item  Without a centralized auditor, the blockchain maintains a publicly auditable transaction record. By querying it, the CHs are able to autonomously learn the market state and adapt their caching strategies accordingly.
 \item The smart contracts use financial incentive to ensure the interests of different parties in a trustless caching market.
 \item The proposed consensus protocol financially incentivizes the CHs to stay active in their service while securing the blockchain consistency with little resource consumption.
\end{itemize}
We model the blockchain-based content caching market as a Chinese restaurant game~\cite{6332480}. Based on the analysis of the Nash Equilibrium (NE) of the game, we propose a decentralized NE algorithm using sequential best response. The simulation results demonstrate the efficiency and reliability of the proposed decentralized NE searching algorithm.

\section{Network Model and Design Approach}\label{Sec_sys_description}
\subsection{Network Structure}
\label{sub_sec_model}
We consider a hierarchical wireless network (e.g., an ultra dense small cell network), where $N$ CPs offer a static, homogeneous catalog of contents, $\mathcal{K}\!=\!\{1, \ldots, K\}$, to mobile users. $M$ edge nodes, i.e., the SCBSs and user-deployed Wi-Fi APs, work as the proactive CHs in the network. We assume that the mobile users are indifferent in choosing the services of any CP, and each user randomly subscribes to one of the CPs with a flat-rate subscription payment for the full access to $\mathcal{K}$. We also consider that the frequency of user demands for the contents remains the same for a sufficiently long period. Then, after arranging the contents in $\mathcal{K}$ in a descending order of their popularity, we can use the Zipf distribution to model the probability of a content being requested during a given time~\cite{749260}:
\begin{equation}
  \label{eq_zpfs}
  p_k=\frac{k^{-\beta}}{\sum_{k=1}^Kk^{-\beta}},
\end{equation}
where the coefficient $\beta$ reflects how skewed the distribution is.

We assume that a CH $m$ chooses to cache only one content from one CP during a certain period, mainly due to the storage/computation limit. In return, the corresponding CP does not pay the CH for merely prefetching contents, but promises to offer reward according to the number of offloaded deliveries made by the CH to the mobile users. We also assume that the price of delivery offloading may differ for each CP. As shown in Figure~\ref{fig_structure}, the transactions occur only from the mobile users to the CPs and from the CPs to the CHs. In this way, the caching process is kept transparent to the mobile users for most of the time, except the final content delivery stage from the CHs to the mobile users.

\subsection{Blockchain for a Distributed Cache-delivery Market}
\label{sub_sec_bc}
\begin{figure}[t]
\centering
\includegraphics[width=.40\textwidth]{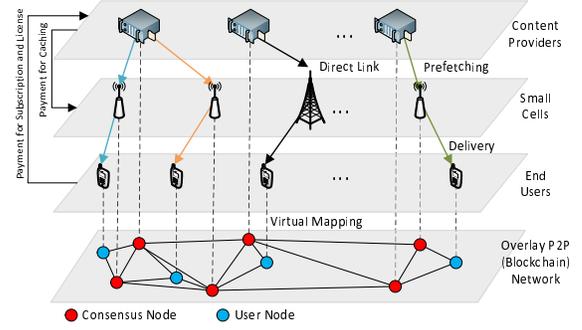}
\caption{Illustration of payment flow and blockchain network structure for proactive caching in a hierarchical wireless network.}\label{fig_structure}
\end{figure}

\begin{figure}%
  \centering
  \tikzstyle{atomic_block_basic}=[draw, fill=gray!10, text width=2.5 cm, text centered, minimum height=1.8em]
  \tikzstyle{atomic_block_other}=[fill=white!10, text width=0.3 cm, text centered, minimum height=1.8em]
  \subfigure[]
  {\label{fig_tx}
    \begin{tikzpicture}[font=\scriptsize, scale=0.75, every node/.style={transform shape}]

        \node [scale=1] (head) at (0, 0) [atomic_block_basic] {Transaction ID (Hashcode): $\mathcal{H}(tx_{i,j})$};
        \node [scale=1] (sender) at (0, -0.75) [atomic_block_basic] {Sender's Address: $\mathcal{H}(pk_{i})$};
        \node [scale=1] (receiver) at (0, -1.5) [atomic_block_basic] {Receiver's Address: $\mathcal{H}(pk_{j})$};
        \node [scale=1] (data) at (0, -2.25) [atomic_block_basic] {Payload Data: Content ID, Payment Value, Transaction Fee, etc.};
        \node [scale=1] (signature) at (0, -3.05) [atomic_block_basic] {Sender's Signature: $\sigma_i$};

    \end{tikzpicture}
  }%
  \qquad
  \subfigure[]
  {\label{fig_blockchain}
    \begin{tikzpicture}[font=\scriptsize, scale=0.75, every node/.style={transform shape}]

        \node [scale=1] (head1) at (0, 0) [atomic_block_basic] {Block Head: $\mathcal{H}(B_{t})$};
        \node [scale=1] (oldhead1) at (0, -0.68) [atomic_block_basic] {Head of Precedent Block: $\mathcal{H}(B_{t-1})$};
        \node [scale=1] (ID1) at (0, -1.36) [atomic_block_basic] {Timestamp (ID): $t$};
        \node [scale=1] (signature1) at (0, -2.02) [atomic_block_basic] {Block Generator's Proof: $(pk_m, \sigma_m)$};
        \node [scale=1] (TX1) at (0, -2.9) [atomic_block_basic] {Set of Transactions (Organized as a Merkle Tree): $\mathcal{TX}_{t}$};

        \node [scale=1] (head2) at (-3.5, 0) [atomic_block_basic] {Block Head: $\mathcal{H}(B_{t-1})$};
        \node [scale=1] (oldhead2) at (-3.5, -0.68) [atomic_block_basic] {Head of Precedent Block: $\mathcal{H}(B_{t-2})$};
        \node [scale=1] (ID2) at (-3.5, -1.36) [atomic_block_basic] {Timestamp (ID): $t\!-\!1$};
        \node [scale=1] (signature2) at (-3.5, -2.02) [atomic_block_basic] {Block Generator's Proof: $(pk_{m'}, \sigma_{m'})$};
        \node [scale=1] (TX2) at (-3.5, -2.9) [atomic_block_basic] {Set of Transactions (Organized as a Merkle Tree): $\mathcal{TX}_{t-1}$};

        \node [scale=1] (others) at (-5.9, -0) [atomic_block_other] {$\ldots$};
        \draw[-latex] (oldhead1.west) -- ++(-0.4,0) |- (head2.east);
        \draw[-latex] (oldhead2.west) -- ++(-0.4,0) |- (others.east);
    \end{tikzpicture}
  }%
  \caption{Data structures for constructing a blockchain. (a) Data fields in a single transaction. (b) Blockchain as a hash linked list of blocks.}%
  \label{fig:fig_data_structure}%
\end{figure}
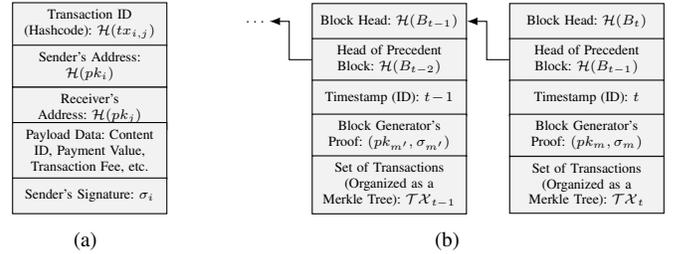

We employ the virtual P2P blockchain network as the backbone of the decentralized content delivery market (see Figure~\ref{fig_structure}). An entity (e.g., a CP, a CH or a mobile user) $i$ in the wireless network maps itself as a node in the blockchain network based on a pair of asymmetric keys, namely, a secret key $sk_i$ and a public key $pk_i$. Let $\mathcal{H}(\cdot)$ denote a collision-resistant, irreversible hash function. Node $i$ identifies itself on the blockchain by a unique ``transaction address'', $\mathcal{H}(pk_i)$, using the hashcode of its public key. We consider that all payments are made in blockchain tokens (c.f., Ethers of the Ethereum network~\cite{wood2014ethereum}). Then, the two types of transactions in the caching market can be abstracted as follows (see also Figure~\ref{fig_tx}):
\begin{Definition}[Transaction]
  \label{def_TX} A transaction $tx_{i,j}$ from node $i$ to node $j$ on the blockchain can be represented by a 4-tuple: $tx_{i,j}=\langle \mathcal{H}(pk_i), \mathcal{H}(pk_j), d_{i,j},  \sigma_i\rangle$, where
  \begin{itemize}
    \item $\mathcal{H}(pk_i)$ and $\mathcal{H}(pk_j)$ are the addresses of nodes $i$ and $j$,
    \item $d_{i,j}$ is the payload data specifying the payment value and the auxiliary information (e.g., the content ID), and
    \item $\sigma_i$ is the publicly verifiable digital signature of node $i$ using the secret key $sk_i$, namely, $\sigma_i=\textrm{Sign}(sk_i, d_{i,j})$.
  \end{itemize}
\end{Definition}

A node issues a new transaction by broadcasting it to the entire blockchain network with an off-the-shelf P2P protocol (e.g., Whisper in Ethereum~\cite{wood2014ethereum}). To further maintain a trustworthy and strictly-ordered log of transactions, the blockchain network records the transactions in a chain of ``blocks'', where each block contains a unique set of transactions. By abstracting away the implementation details, we define the data structure of the block and the blockchain as follows (see also Figure~\ref{fig_blockchain}):
\begin{Definition}[Block]
  \label{def_block} A block $B_t$ can be represented by a 5-tuple: $B_t=\langle t, \mathcal{TX}_t, (pk_m, \sigma_m), H(B_t), H(B_{t-1}) \rangle$, where
  \begin{itemize}
    \item $t$ is the ID (i.e., timestamp) of block $B_t$ in the blockchain.
    \item $\mathcal{TX}_t=\{tx_1,\ldots, tx_l\}$ is a set of unique transactions that do not conflict with the transactions in $B_0$ to $B_{t-1}$.
    \item $(pk_m, \sigma_m)$ contains the public key and the signature of node $m$ that publishes the block. $\sigma_m=\textrm{Sign}(sk_m, \mathcal{TX}_t)$.
    \item ${H}(B_t)$ and ${H}(B_{t-1})$ are the heads (i.e., hashcodes) of the current block and the precedent block at $t\!-\!1$. ${H}(B_t)=\mathcal{H}(t\Vert\mathcal{TX}_t\Vert (pk_m,\sigma_m)\Vert H(B_{t-1}))$.
  \end{itemize}
\end{Definition}

\begin{Definition}[Blockchain]
  \label{def_blockchain}
  A blockchain $\mathcal{C}(t)\!=\!\{B_0, \ldots, B_t\}$ is a sequence (i.e., linked list) of blocks indexed in a strictly increasing time order. $B_0$ is an arbitrary block known as the genesis block of $\mathcal{C}(t)$, and $B_t$ is known as the head of $\mathcal{C}(t)$.
\end{Definition}

With a chain of blocks linked by the block heads (i.e., hashcodes), a node have to modify every subsequent block if it wants to forge any transaction in a block stored at the local replica of the blockchain. Therefore, if the network adopts a reliable consensus mechanism to coordinate the states of the local replicas of the blockchain, the blockchain works as a tamper-proof public database of the transactions in the content delivery market~\cite{7423672}. Then, each CH can easily count the frequency of requests for a certain content $k\!\in\!\mathcal{K}$ by incrementally querying the blockchain about the related new transactions sent by the mobile users. The CHs may use the observed request frequencies to estimate the content popularity distribution in (\ref{eq_zpfs}), and thus are able to autonomously choose the contents to cach according to the demand state in the market.

We consider that the CHs and CPs work as the full (consensus) nodes and participate in the consensus process of the blockchain network. The mobile devices work as lightweight nodes and only issue requests for content delivery. Instead of relying on the computation-intensive protocol using Proof of Work (PoW)~\cite{7423672}, we adopt a PoS-based consensus mechanism from the Ouroboros protocol~\cite{Kiayias2017}. Without loss of generality, we assume that the consensus nodes are equipped with roughly synchronized clocks. A new block is generated in a slotted manner with a fixed time interval. The PoS scheme facilitates an uncompromisable random leader-election process at each time slot to designate a unique consensus node for block generation. To ensure the sustainability of the proactive caching eco-system, we make the following assumptions regarding the token supply mechanism of the blockchain:
\begin{itemize}
  \item Each CP is assigned a sufficient amount of tokens at $B_0$.
  \item New tokens are supplied to the blockchain only through a fixed-amount reward for generating every new block.
  \item The CPs provide a fixed exchange rate between the tokens and the fiat money for the CHs and the mobile users.
  \item The transaction fee (c.f., \emph{gas} in Ethereum~\cite{wood2014ethereum}) is negligible.
\end{itemize}
From~\cite{Kiayias2017}, we introduce the formal definition of the leader-election process for block generation with PoS as follows.
\begin{Definition}
  \label{def_leader_selection}
  Assume that at time slot $s_t$ the stakes held by the $M\!+\!N$ consensus nodes are static and measured by $\mathbf{u}(s_t)\!=\![ u_1(s_t),\ldots,u_{M+N}(s_t)]^{\top}$. A leader-election process consists of a distribution $\mathcal{D}$ and a deterministic function $F(\cdot)$ such that, with a random seed $\rho\!\leftarrow\!\mathcal{D}$, $F(\mathbf{u}, \rho, s_t)$ outputs a unique leader index $m$ ($1\!\le\!m\!\le\!M\!+\!N$) for block generation with probability
\begin{equation}
  \label{eq_leader_selection}
  p^{win}_{m}(s_t)=\frac{u_m(s_t)}{\sum_{i=1}^{M+N}u_i(s_t)}.
\end{equation}
The random variable $m\!\leftarrow\!F(\mathcal{S}, \rho, s_t)$ is independent of $s_t$.
\end{Definition}

The PoS-based election scheme can be implemented by a standard Follow-the-Satoshi (FtS) algorithm~\cite{Kiayias2017}. The FtS resembles the process of biased coin-tossing and randomly selects a leader's address. This is achieved through indexing a sub-set of tokens controlled by the consensus nodes (i.e, stakeholders) and tracking the owner's address of a random token index\footnote{FtS can be performed in advance at $s_t\!-\!1$ by a smart contract through executing a random search in the Merkle tree of the related stake deposits. See \url{https://github.com/Realiserad/fts-tree} for a simplified implementation.}. The randomness of token selection is guaranteed by the seeding function $\rho\!\leftarrow\!\mathcal{D}$, which is implemented using the hashcode of the precedent block head, i.e., $\sigma\!\leftarrow\!\mathcal{H}(B_{s_t-1})$. Here, $\mathcal{H}(\cdot)$ is treated as a trusted uniform random oracle~\cite{Kiayias2017} and $\sigma$ is used to determine the random token index (e.g., with a simple modulo operation). When all the consensus nodes are honest, the proposed PoS scheme is able to avoid the inherent inefficiency of PoW due to ``block mining'' competition~\cite{7423672} such as computational resource consumption and block orphaning.

Let an \emph{epoch} define a fixed time interval consisting of $T$ time slots. Instead of using the balance of each stakeholder as its stake in (\ref{eq_leader_selection}) (c.f.,~\cite{Kiayias2017}), we measure the stake of each CH by its delivery reward collected in the latest epoch. Such design incentivizes the CHs to stay active online, since a CH no longer receives any profit by only holding the tokens. On the other hand, we consider that the CPs' stakes are proportional to the amount of unconfirmed payment for content delivery in the last epoch. When the CHs fail to complete the delivery tasks or when pending transactions accumulate due to the Denial of Service (DoS) attacks by malicious CHs, such design ensures the CPs to take control of the block generation process with high probability. Thereby, the blockchain lays the foundation for implementing a secured decentralized caching market.

\subsection{Caching and Delivery Based on Smart Contracts}
\begin{figure}[t]
\centering
\includegraphics[width=.345\textwidth]{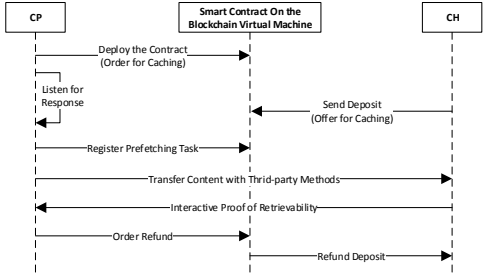}
\caption{Sequence diagram of the content prefetching contract.}\label{fig_uml_contract_prefetch}
\end{figure}

Now, we address the issues of autonomous content delivery and truthfulness enforcement with smart contracts~\cite{wood2014ethereum}. From Section~\ref{Sec_sys_description}, we identify two stages, namely, the prefetching and the delivering stages, in the content delivery process and implement for each stage a group of smart contracts~\cite{Solidity}. At the prefetching stage, the smart contract is used by the CPs and CHs for negotiating about the assignment of offloaded contents. As shown in Figure~\ref{fig_uml_contract_prefetch}, a CP $n$ posts for each content $k\!\in\!\mathcal{K}$ an \emph{order for caching} by deploying a corresponding smart contract, which specifies the price, $o_{n,k}$, for one successfully offloaded delivery in the future. An interested CH $m$ may respond to CP $n$ by calling the function \emph{offer for caching} for content $k$ and sending a deposit to the smart contract. Consequently, an event \emph{response} is fired by the smart contract to notify CP $n$ about the response from CH $m$. We consider that a CP is able to register more than one CHs for the delivery task (see the ``crowdfund contract'' in Ethereum~\cite{Solidity} for example). CP $n$ confirms to choose CH $m$ by calling a registering function of the smart contract. Then, it transfers a copy of content $k$ to CH $m$ with third-party methods. To get refunded, CH $m$ has to provide CP $n$ an interactive Proof of Retrievability (PoR)~\cite{7966965} for content $k$ in the form of a series of Merkle proofs~\cite{7966965}. After verifying the PoR, CP $n$ orders to refund the deposit to CH $m$. At the end of each epoch, the existing contracts are destroyed by the CPs to avoid unresolved transactions. A set of new contracts in the same form are deployed at the beginning of the next epoch.

\begin{figure}[t]
\centering
\includegraphics[width=.37\textwidth]{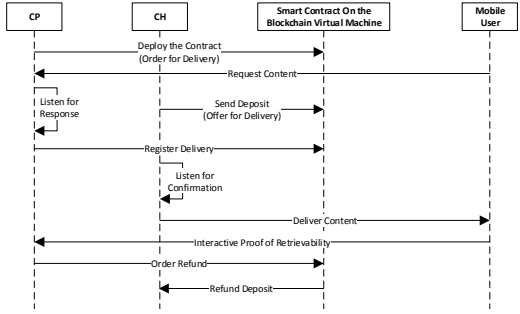}
\caption{Sequence diagram of the content delivery contract.}\label{fig_uml_contract_delivery}
\end{figure}

At the delivering stage, CP $n$ posts for each content $k\!\in\!\mathcal{K}$ an \emph{order for delivery} by deploying a smart contract, which works as an escrow account to ensure that the payment by CP $n$ is made only when the delivery is successful. When a CH $m$ responds by calling the function  \emph{offer for delivery}, it sends a security deposit to the smart contract. CP $n$ evenly distribute the delivery tasks among the responding CHs and notifies the smart contract to register each CH $m$ for the delivery task. On the other hand, the payment of CP $n$ for the delivery, $o_{n,k}$, is also held by the smart contract until the target mobile user provides the interactive PoR of content $k$ to CP $n$. To prevent the CHs and the flat-rate mobile users jointly cheating for the delivery reward without actually delivering the content, the smart contracts require that the mobile users provide the PoR within a certain delay to confirm the content reception. Otherwise, the contract will rollback and CP $n$ will receive CH $m$'s security deposit as a punishment. Such design ensures that without a locally stored content copy, a mobile user is not able to obtain the PoR from the CH due to transmission delay. Finally, the content delivery contract is also updated at the beginning of each epoch to remove the unfinished transactions.

\section{Blockchain-Based Content Caching Game}
\label{sec_game_analysis}
\subsection{Caching Strategies with PoS-Based Block Generation}
\label{sub_sec_strategy}
We consider that during each epoch, a total number of $L$ users request content delivery in the network. Since the users are indifferent in subscribing to any CP for the same content service, from (\ref{eq_zpfs}), the expected number of requests for content $k$ in CP $n$ during the epoch can be written as $L(n,k)\!=\!LN^{-1}p_k$. Assume that a subset of CHs, $\mathcal{M}_{n,k}$, have cached the ready-to-deliver copies of content $k$ provided by CP $n$. Again, if each CP evenly distributes the delivery tasks among the responding CHs, the expected number of delivery tasks assigned to CH $m\!\in\!\mathcal{M}_{n,k}$ can be expressed as $L_m(n,k)\!=\!LN^{-1}M_{n,k}^{-1}p_k$, where $M_{n,k}$ is the size of $\mathcal{M}_{n,k}$ and $\sum_{n=1}^{N}\sum_{k=1}^{K}{M}_{n,k}\!=\!M$. By adopting the strategy of best-effort delivery, CH $m$ offloads all the delivery tasks assigned by CP $n$. Then, for a given sub-set of CHs caching content $k$ for CP $n$,  $\mathcal{M}_{n,k}$, the expected delivery reward received by CH $m$ ($m\!\in\!\mathcal{M}_{n,k}$) can be expressed as:
\begin{equation}
  \label{eq_best_effort_reward}
  r^d_m(n,k)=o_{n,k}L_m(n,k)=o_{n,k}LN^{-1}M_{n,k}^{-1}p_k.
\end{equation}

If CH $m$ is selected as the block generation leader at one epoch, it also receives a fixed reward for generating $T$ new blocks. Let $\lambda$ denote the total block generation reward in an epoch. According to Section~\ref{sub_sec_bc}, the probability for CH $m$ to be elected in the next epoch is determined by the ratio between its own delivery reward and the sum of the delivery reward collected by all the CHs in the current epoch. By (\ref{eq_leader_selection}) and (\ref{eq_best_effort_reward}), for CH $m$ caching content $k$ from CP $n$, the expected block generation reward to be collected in the next epoch is
\begin{equation}
  \label{eq_expected_mining_reward}
  r^m_m(n,k)=\lambda \frac{o_{n,k}L_m(n,k)}{\sum_{i=1}^N\sum_{j=1}^K\sum_{l\in\mathcal{M}(i,j)}o_{i,j}L_l(i,j)},
\end{equation}
where $L_l(i,j)$ is the expected number of tasks assigned to CH $l$ for delivering content $j$ for CP $i$.

Now, we can define the action of CH $m$ in an epoch as the pair of the selected CP and content, i.e., $a_m=(n,k)$. We note that the expected reward provided by CP $n$ for delivering content $k$ is evenly distributed to the set of corresponding CHs, $\mathcal{M}_{n,k}$. Then, the payoff for CH $m$ to take action $a_m=(n,k)$ can be obtained by adding up (\ref{eq_best_effort_reward}) and (\ref{eq_expected_mining_reward}) as follows:
\begin{equation}
  \label{eq_rational_mining_reward}
  r_m(a_m)= \frac{o_{n,k}Lp_k}{NM_{n,k}}+\lambda \frac{o_{n,k}p_k}{M_{n,k}\sum_{i=1}^N\sum_{j=1}^K o_{i,j}p_j},
\end{equation}
where owning to the adjusted stakes based on the unspent reward held by the CPs, $\sum_{i=1}^N\sum_{j=1}^K o_{i,j}p_j$ is independent of $M_{i,j}$. Following the proposed mechanism of stake evaluation, the payoff of a CH in (\ref{eq_rational_mining_reward}) only depends on the joint action of the CHs at the current epoch. Thereby, we are ready to investigate the caching strategy selection of the CHs through modeling the caching market as a non-cooperative game.

\subsection{Caching Market as a Chinese Restaurant Game}
\label{sub_sec_sub_game_formulation}
Let $\mathcal{A}\!=\!\{(n,k)\!:\!1\!\le\!n\!\le\!N, 1\!\le\!k\!\le\!K\}$ denote the set of actions (i.e., CP-content choices) for CH $m\!\in\!\mathcal{M}\!=\!\{m\!:\!1\!\le\!m\!\le\!M\}$, and $\mathbf{a}\!=\![a_1,\ldots,a_M]^{\top}$ denote the vector of the CHs' joint actions. We can define the non-cooperative caching game among the CHs as a 3-tuple: $\mathcal{G}\!=\!\langle \mathcal{M}, \mathcal{A}, \{r_m(\mathbf{a})\}_{m\in\mathcal{M},a_m\in\mathcal{A}}\rangle$. We define the \emph{grouping of actions} as the vector of the numbers of the CHs choosing the same action given a joint action $\mathbf{a}$: $\mathbf{g}_{\mathcal{M}}(\mathbf{a})\!=\![M_{1,1},\ldots, M_{a},\ldots,M_{N,K}]^{\top}$. From (\ref{eq_rational_mining_reward}), we define the \emph{reward potential} of each action as ${R}_{n,k}\!=\!o_{n,k}p_k$. Then, for CH $m$ choosing $a\!=\!(n,k)$, when fixing the other CHs' actions, we can rewrite (\ref{eq_rational_mining_reward}) as:
\begin{equation}
  \label{eq_modified_mining_reward}
  r_m(a)\!=\!r(R_{a}, M_{a})\!=\!
  \frac{L}{N}\frac{R_{a}}{M_{a}}\!+\!
  \frac{\lambda }{\sum_{i=1}^N\sum_{j=1}^K R_{i,j}}\frac{R_{a}}{M_{a}}.
\end{equation}
Since $r(R_{a}, M_{a})$ is a decreasing function of $M_a$, $M_{a}$ reflects the impact of the negative network effect on CH $m$'s payoff for choosing action $a$. By inspecting the first order derivative of $r(R_{a}, M_{a})$ with respect to $R_a$, we can easily show that $r(R_{a}, M_{a})$ is an increasing function of $R_{a}$. Then, game $\mathcal{G}$ is a typical Chinese restaurant game~\cite{6332480}, where $\mathcal{A}$ can be compared to the table set in a restaurant, and $R_a$ and $M_{a}$ can be compared to the size and the customer number of table $a$, respectively.

Let $a_{-m}$ denote the adversaries' actions with respect to CH $m$.
Let $\mathbf{g}_{\mathcal{M}}(a_{-m})\!=\![M^{-m}_{1,1},\ldots,M^{-m}_{N,K}]^{\top}$ denote the outcome of action grouping for ${a}_{-m}$, where
$M^{-m}_{n,k}$ is the number of CHs except CH $m$ choosing action $(n,k)$. Then, we can define the Best Response (BR) of CH $m$ to $a_{-m}$ as:
\begin{equation}
  \label{eq_br}
  BR_m(a_{-m})\!=\!BR_m\left(\mathbf{g}_{\mathcal{M}}(a_{-m})\right)\!=\!\arg\max_{a\in\mathcal{A}}r(R_{a}, M^{-m}_{a}\!+\!1).
\end{equation}
Based on (\ref{eq_br}), we can define the NE of game $\mathcal{G}$ in the form of the simultaneous best response as follows.
\begin{Definition}
  \label{def_NE}
  A joint CH action $\mathbf{a}^*=(a^*_m,a^*_{-m})$ is an NE of game $\mathcal{G}$, if $\forall m\in\mathcal{M}$, the following holds:
  \begin{equation}
    \label{eq_ne}
    a^*_m\!=\!BR_m\left(\mathbf{g}_{\mathcal{M}}^{*}(a_{-m})\right)\!=\!\arg\max_{a\in\mathcal{A}}r(R_{a}, M^{*,-m}_{a}\!+\!1),
  \end{equation}
  where $\mathbf{g}_{\mathcal{M}}^{*}(a_{-m})$ is the outcome of action grouping for $a^*_{-m}$.
\end{Definition}

\subsection{Nash Equilibrium of the Caching Game}
\label{sub_sec_leader_analysis}
After reformulating the payoff of a CH $m$ with action $a=(n,k)$ as a function of $R_a$ and $M_a$ in (\ref{eq_modified_mining_reward}), we note that the CHs choosing the same action receive an identical payoff. Then, we can show that game $\mathcal{G}$ is an exact potential game.
\begin{Theorem}
 \label{thm_potential_game}
 $\mathcal{G}$ is an exact potential game and admits at least one pure-strategy NE.
\end{Theorem}
\begin{proof}
 We define the following potential function (i.e., Rosenthal's potential function~\cite{han2012game}) of $\mathbf{a}\!=\!(a_m,a_{-m})$ in game $\mathcal{G}$:
 \begin{equation}
  \label{eq_potential_function}
\phi(\mathbf{a})=\sum_{a\in\mathcal{A}}\sum_{i=1}^{M_{a}}r(R_a,i),
 \end{equation}
where $M_{a}$ is given by the action grouping $\mathbf{g}_{\mathcal{M}}$ of $\mathbf{a}\!=\!(a_m\!=\!a,a_{-m})$. If CH $m$ unilaterally switches its action from $a_m\!=\!(n,k)$ to $a'_m\!=\!(n',k')$, the change only affects the CHs that adopt actions $a_m$ and $a'_m$. Thus, we obtain a new grouping of actions $\mathbf{g}_{\mathcal{M}}(\mathbf{a}')\!=\![M'_{1,1},\ldots, M'_{N,K}]^{\top}$, where the only difference from $\mathbf{g}_{\mathcal{M}}(\mathbf{a})$ is that $M_{a_m}\!=\!M'_{a_m}+1$ and $M_{a'_m}\!=\!M'_{a'_m}\!-\!1$. Then, we obtain
\begin{equation}
 \label{eq_proof}
 \begin{array}{ll}
  \phi(a_m,a_{-m})-\phi(a'_m, a_{-m})\\
  =\bigg(\sum\limits_{i=1}^{M_{a_m}}r(R_{a_m},i)+\sum\limits_{i=1}^{M_{a'_m}}r(R_{a'_m},i)\bigg)\!-\!
  \bigg(\sum\limits_{i=1}^{M_{a_m}\!-\!1}r(R_{a_m},i)\\
  +\!\sum\limits_{i=1}^{M_{a'_m}+1}r(R_{a'_m},i)\bigg)=r(R_{a_m},M_{a_m})-r(R_{a'_m},M_{a'_m}\!+\!1)\\
  =r_m(a_m,a_{-m})-r_m(a'_m,a_{-m}).
 \end{array}
\end{equation}
By Definition 52 in~\cite{han2012game}, game $\mathcal{G}$ is an exact potential game. Then, Theorem~\ref{thm_potential_game} immediately follows Theorem 53 in~\cite{han2012game}.
\end{proof}
\begin{Corollary}
 \label{co_convergence}
 For game $\mathcal{G}$, the asynchronous/sequential best-response dynamics converge with probability one to a pure NE.
\end{Corollary}
\begin{proof}
 Given Theorem~\ref{thm_potential_game}, namely, $\mathcal{G}$ is an exact potential game, Corollary~\ref{co_convergence} follows Theorem 143 in~\cite{han2012game}.
\end{proof}
Considering the asynchronous nature of the caching choice selection based on the blockchain, we propose in Algorithm~\ref{alg1} the asynchronous best-response algorithm for equilibrium strategy searching in the content market game $\mathcal{G}$. By Corollary~\ref{co_convergence}, the proposed strategy searching scheme in Algorithm~\ref{alg1} is guaranteed to converge to a pure-strategy NE.
\begin{algorithm}[t]
  \caption{Equilibrium Strategy Searching}
 \begin{algorithmic}[1]
  \REQUIRE
  Randomly initialize $\mathbf{a}(0)=[a_1(0),\ldots,a_m(0)]^{\top}$;
  \WHILE {$\mathbf{a}(t)$ not converged}
  \FOR{CH $m\in\mathcal{M}$}
  \STATE $r_m\leftarrow0$;
    \FOR {$a\in\mathcal{A}$}
      \IF{$r(R_a,M^{-m}_a+1) > r_m$}
	\STATE $a_m\leftarrow a$, $r_m\leftarrow r(R_a,M^{-m}_a+1)$;
      \ENDIF
    \ENDFOR
    \STATE $a_m(t)\leftarrow a_m$;
  \ENDFOR
  \STATE $t\leftarrow t+1$;
  \ENDWHILE
 \end{algorithmic}
  \label{alg1}
\end{algorithm}

\section{Simulation Results}\label{Sec:Simulation}
\begin{figure}[t]
\centering     
\subfigure[]{\label{performance_comparison}\includegraphics[width=.235\textwidth]{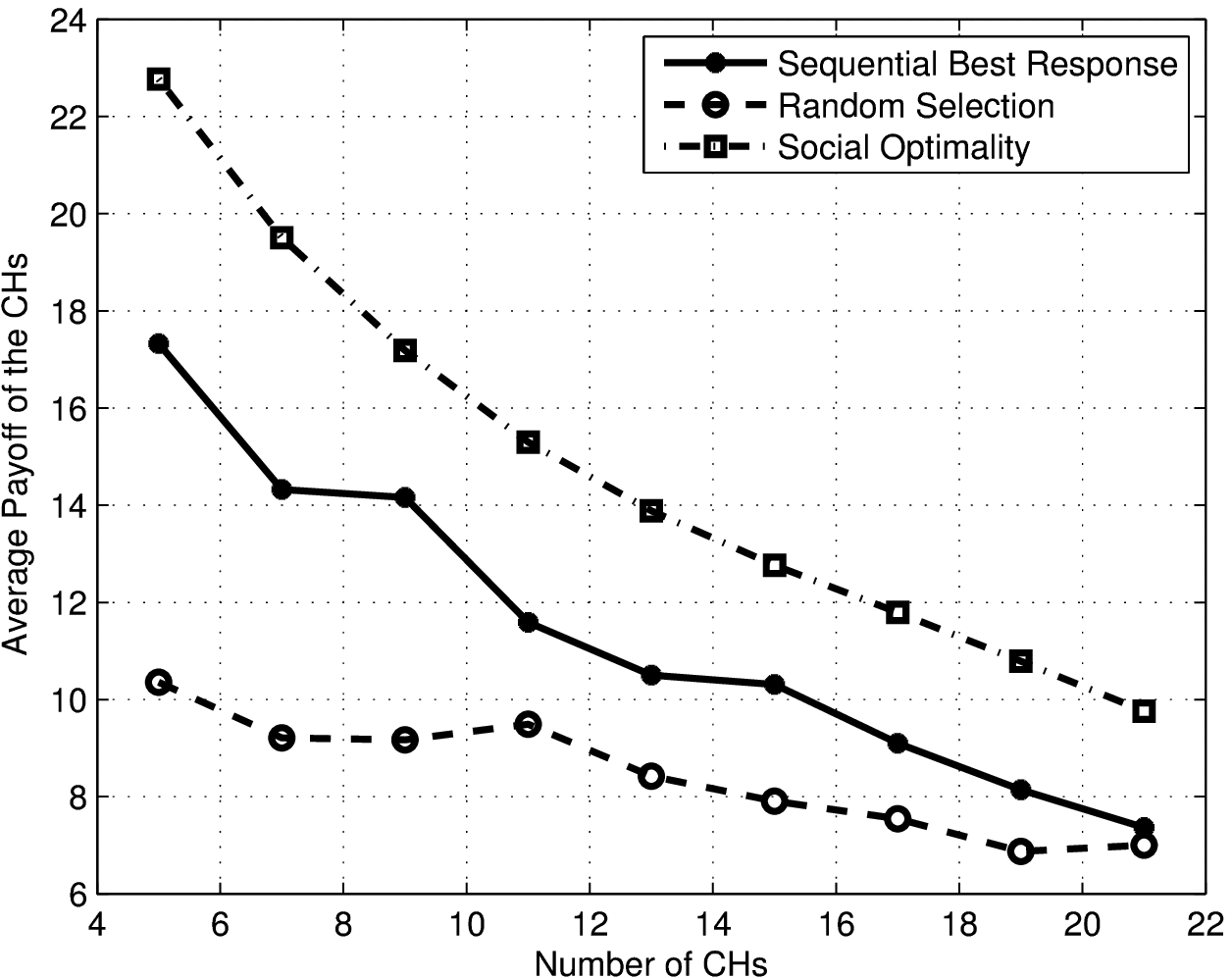}}
\subfigure[]{\label{performance_comparison_num_delivery}\includegraphics[width=.24\textwidth]{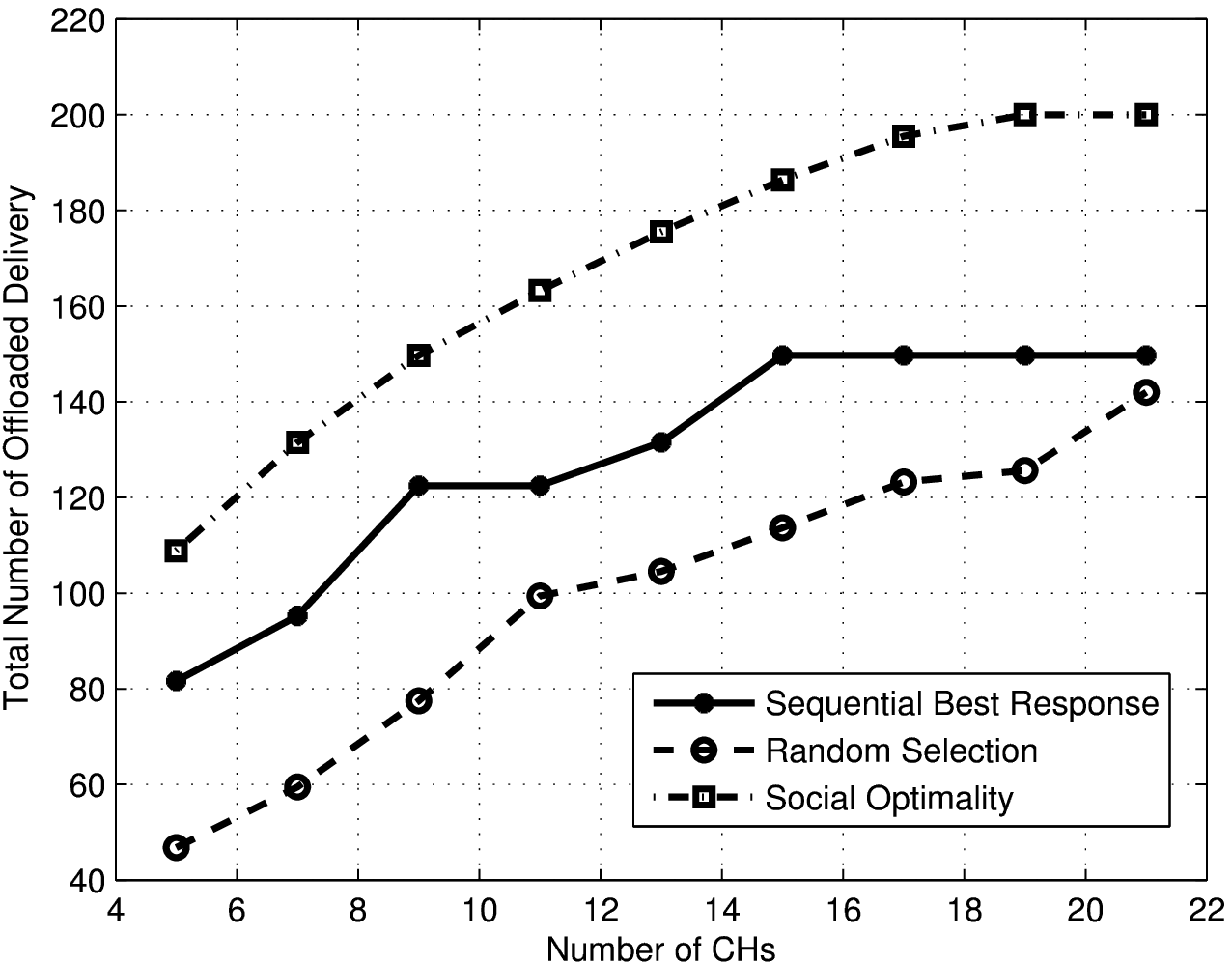}}
\caption{(a) Average payoff of the CHs vs. number of CHs in the network. (b) Number of offloaded content deliveries to the CHs vs. number of CHs.}
\label{fig_sim_performance}
\end{figure}

We first consider a network of $N\!=\!3$ CPs with a catalog of $K\!=\!6$ contents. There are $L\!=\!200$ mobile users in the network and the skewness of the popularity distribution in (\ref{eq_zpfs}) is set to $\beta\!=\!1$. We first consider that the delivery rewards are identical for all the CP-content pairs. In Figure~\ref{fig_sim_performance}, we compare the performance of the proposed best-response strategy searching scheme in Algorithm~\ref{alg1} with that of the random content selection and that of the centralized payoff optimization, respectively. The result for random content selection is obtained with Monte Carlo simulation. As we can observe in Figures~\ref{performance_comparison} and~\ref{performance_comparison_num_delivery}, the performance of the proposed caching strategy searching algorithm significantly outperforms that of random content selection in both the CHs' average payoff and the total number of offloaded deliveries. However, the gap of performance between the centralized optimum and the proposed algorithm indicates that the ``pure price of anarchy'' is unnegligable for the caching game $\mathcal{G}$. Theoretically, from the individual payoff function given by (\ref{eq_modified_mining_reward}), we can re-interpret the caching game $\mathcal{G}$ as a singleton congestion game~\cite{han2012game}. Then, without coordination, the CHs may be reluctant to offer caching services to the CP-content pairs that pay less for delivery, although by doing so they are able to increase the social welfare of the CHs. Instead, sharing the delivery demands with other CHs for the CP-content pairs with higher reward potential will lead to a higher individual payoff.

\begin{figure}[t]
\centering     
\subfigure[]{\label{action_grouping}\includegraphics[width=.235\textwidth]{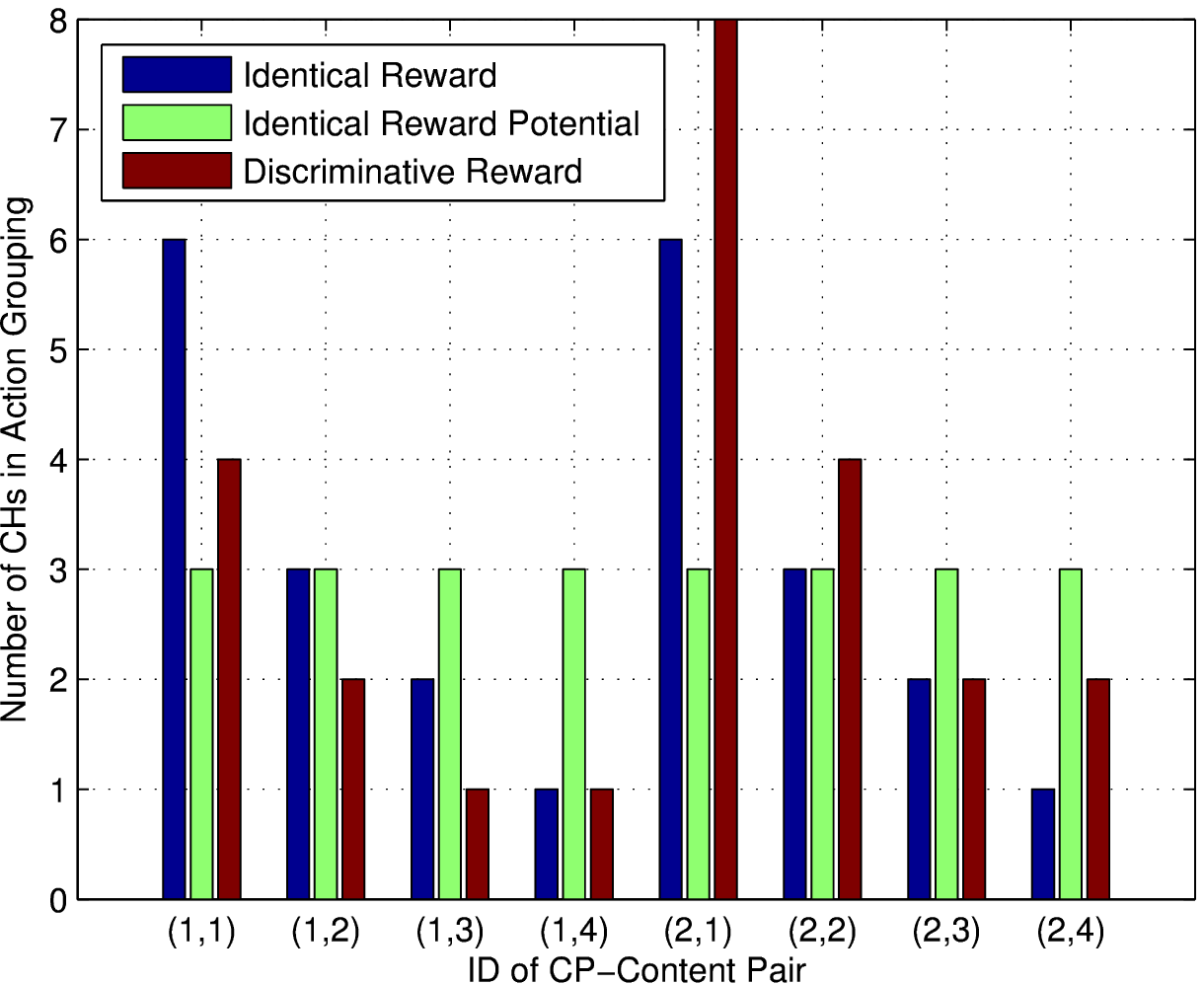}}
\subfigure[]{\label{payoff_per_CH}\includegraphics[width=.24\textwidth]{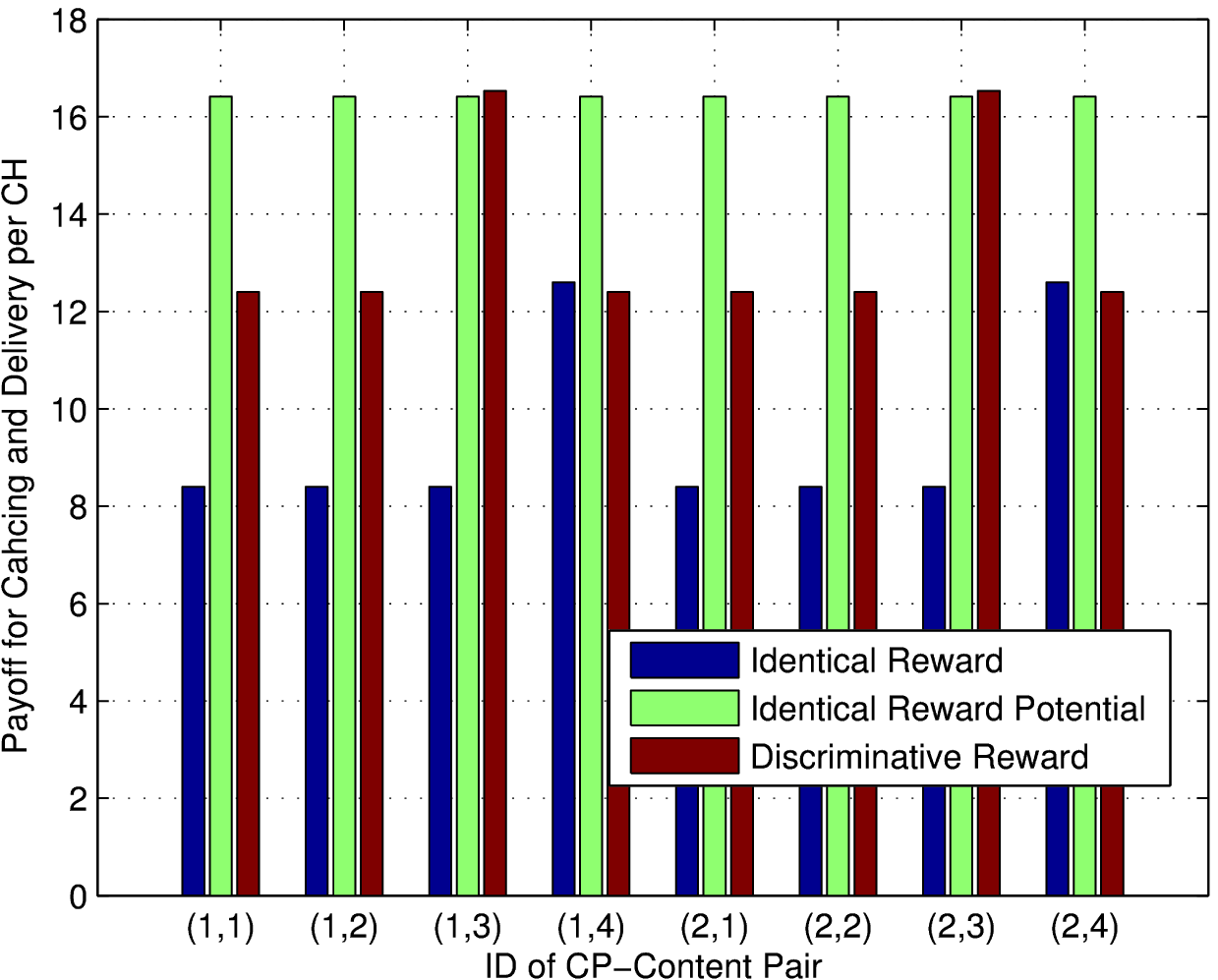}}
\caption{(a) Action grouping results with different rewarding schemes. (b) Payoff comparison of action groups with different rewarding schemes.}
\label{fig_sim_action_grouping}
\end{figure}

In Figure~\ref{fig_sim_action_grouping}, we study the impact of the CPs' rewarding scheme on the CHs' NE strategies. For ease of exposition, we consider a network of 2 CPs with 4 contents, 24 CHs and 200 mobile users. We consider 3 pricing schemes: (a) the CPs offering uniform reward for each content, i.e., $o_{n,k}\!=\!o_{n',k'}$ for all feasible $n,n',k,k'$ (the blue bars in Figure~\ref{fig_sim_action_grouping}); (b) the CPs offering different reward to make the reward potential identical, namely, $o_{n,k}p_k\!=\!o_{n',k'}p_{k'}$ for all feasible $n,n',k,k'$ (the green bars in Figure~\ref{fig_sim_action_grouping}); (c) discriminative reward with $o_{2,k}\!=\!2o_{1,k}$ for all feasible $k$ (the red bars in Figure~\ref{fig_sim_action_grouping}). As shown in Figure~\ref{fig_sim_action_grouping}, the reward pricing scheme with identical reward potential is able to achieve better fairness among the CHs, and the CHs are more willing to serve the caching and delivery tasks for the less popular contents. On the other hand, by increasing the level of delivery reward, a CP is able to attract more CHs for its own traffic offloading (see the red bars in Figures~\ref{action_grouping}). We also note in Figure~\ref{payoff_per_CH} that when one CP increases its delivery reward, the payoffs of the CHs at the NE will ultimately increase. This indicates that the CPs are at a higher cost if one of them unilaterally tries to improve its delivery efficiency.

\section{Conclusion}\label{Sec:Conclusion}
We have designed an decentralized proactive caching system in a hierarchical wireless network based on blockchains, which enables the cache helpers to autonomously adapt their strategies without a centralized auditing body. We have designed several smart contacts to enable autonomous caching-delivery task assignment and enforce truthfulness of different parties in the network. We have designed an incentive-compatible consensus mechanism for the blockchain based on proof-of-stake to encourage the cache helpers to stay active in service online. We have modeled the caching system as a Chinese restaurant game and further shown that it is an exact potential game. We have proposed a decentralized strategy searching algorithm based on asynchronous best response for the cache helpers to learn their NE strategies. The numerical simulation results have demonstrated both the efficiency   and the reliability of the proposed proactive caching scheme.

\section{Acknowledgement}
This work was supported in part by Singapore MOE Tier 1 under Grant 2017-T1-002-007, MOE Tier 2 under Grant MOE2014-T2-2-015 ARC4/15, NRF2015-NRF-ISF001-2277 and ISF-NRF grant number 2277/16.

\linespread{1.27}
\bibliographystyle{IEEEtran}
\bibliography{bibfile}

\end{document}